\documentclass[12pt]{article}
\usepackage{amsmath,amssymb,graphicx}
\usepackage{color}
\usepackage{epsfig}
\usepackage{amssymb}
\usepackage{float}
\usepackage{extarrows}
\newcommand{\be}{\begin{equation}}
\newcommand{\bea}{\begin{eqnarray}}
\newcommand{\eea}{\end{eqnarray}}
\newcommand{\ba}{\begin{array}}
\newcommand{\ea}{\end{array}}
\newcommand{\ee}{\end{equation}}
\newcommand{\ml}{\mathcal}

\newcommand{\no}{\nonumber}
\newcommand{\dg}{\dagger}
\newcommand{\de}{\delta}

\begin{document}
\begin{titlepage}

\title{\textbf {Notes on Integrable Boundary Interactions of Open $SU(4)$ Alternating Spin Chains }}
\author{
Jun-Bao Wu$^{a, b}$\footnote{junbao.wu@tju.edu.cn}}
\date{}

\maketitle
\underline{}
\vspace{-10mm}

\begin{center}

{\it
$^{a}$ Department of Physics, Tianjin University, Tianjin 300350, P.~R.~China\\
$^{b}$Center for High Energy Physics, Peking University, Beijing 100871, P.~R.~China}
\vspace{10mm}
\end{center}

\begin{abstract}
\cite{Bai} showed that the planar flavored   Ahanory--Bergman--Jafferis--Maldacena (ABJM) theory is integrable in the scalar sector at two-loop order using coordinate Bethe ansatz.
A salient feature of this case is that the boundary reflection matrices are anti-diagonal with respect to  the chosen basis. In this paper, we relax the coefficients of the boundary terms to be general constants to search for integrable systems among this class. We found that  the only integrable boundary interaction at each end of the spin chain aside from the one in \cite{Bai} is the one with vanishing boundary interactions leading to diagonal reflection matrices. We also construct non-supersymmetric planar flavored ABJM theory which leads to trivial boundary interactions at both ends of the open chain from the two-loop anomalous dimension matrix in the scalar sector.\\


\end{abstract}

\end{titlepage}

\section{Introduction}
Integrability in both gauge theory and string theory sides plays an important role in the study of dynamics of certain AdS/CFT dualities in the planar limit \cite{Beisert:2010jr}. Two canonical examples of such gauge theories are the four-dimensional ${\cal N}=4$ super Yang--Mills theory \cite{Brink:1976bc} and the three-dimensional ${\cal N}=6$  super Chern--Simons theory, which is also known as Ahanory--Bergman--Jafferis--Maldacena (ABJM) theory \cite{ABJM}. The matter fields in the former theory are in the adjoint representation of the gauge group,  while the matter fields in the latter theory are in the bi-fundamental representation of the gauge group.
The first hint for integrability in the planar ${\cal N}=4$ SYM and planar ABJM theories was provided by the fact that the anomalous dimension matrix (ADM) for the single trace operators in the scalar sector at the lowest non-trivial  order in perturbation theory  is given by an integrable Hamiltonian acting on certain closed spin chains \cite{Minahan:2002, Minahan:2008, Bak:2008}. In planar ${\cal N}=4$ SYM theory, such a  closed spin chain has $SO(6)$ symmetry with spin on each site in the fundamental representation. Meanwhile, for planar ABJM theory such a closed spin chain is an alternating one with spin on  even(odd) site in the (anti-)fundamental representation of the symmetry group $SU(4)$.

 With possible applications  to QCD and quantum Hall effects in mind, adding flavors (i. e., matter fields in the  (anti-)fundamental representation) to these AdS/CFT dualities is valuable \cite{Karch:2002sh,Hohenegger:2009, Gaiotto:2009, Hikida:2009}.
When we take the planar (large $N_c$) limit, two different kinds of limits can be considered after flavors are included. The first one is the 't Hooft limit in which one keeps the number of flavors, $N_f$, finite when one lets $N_c$ tend to infinity. In this limit, the contribution from the diagrams involving fundamental matter loops will be suppressed by a positive power of $N_f/N_c$. Another limit one can consider is the Veneziano limit, in which one takes both $N_c, N_f\to \infty$ with their ratio kept finite. In this case, the planar diagrams with fundamental matter loops should be kept. The holographic description of adding flavors in the 't Hooft limit is usually in terms of probe D-branes \cite{Karch:2002sh,Hohenegger:2009, Gaiotto:2009, Hikida:2009}, while in the Veneziano limit the back reaction of the probe branes should be considered to provide  new  supergravity solutions \cite{Burrington:2004id, Kirsch:2005uy, Conde:2011sw}.

After including the fundamental matters, one can also consider the composition operator with (anti-)fundamental fields at two ends and the adjoint/bi-fundament fields in the middle part. This operator naturally maps to state on an open spin chain. A natural question is to determine  whether theories with flavors can lead to an integrable open spin chain in the 't Hooft limit. This question was answered in four-dimensional cases more than a decade ago with  a positive answer, beginning with the results in \cite{Chen:2004mu, Chen:2004yf, Erler:2005nr}. Positive results  for the three-dimensional case were recently provided in \cite{Bai} using the coordinate Bethe ansatz in the case when the obtained flavored ABJM theory has three-dimensional ${\cal N}=3$ supersymmetries\footnote{Strong evidence was quite recently given for the claim that the ${\cal N}=3$ flavored ABJM theory and the ${\cal N}=2$ super Yang-Mills theory with flavors are not integrable in the Veneziano limit \cite{Giataganas:2017guj}.}. The situation in the three-dimensional case is much more complicated than that in the four-dimensional one. The reflection matrices in the three-dimensional case are anti-diagonal with respect to the chosen basis, while those in the four-dimensional cases are diagonal for the cases where the reflection matrices were computed \cite{Erler:2005nr}.

This new feature leads us to think about whether new integrable open spin chains with the same bulk interactions and different boundary interactions exist. We would like to investigate whether such integrable open spin chains with more complicated reflection matrices exist. Too many terms will appear if one considers all of the possible nearest and next-to-nearest neighbor interactions; hence, we focus herein on the interactions studied in \cite{Bai} and consider changing  the coefficients. Within this class of interactions, we find that integrability provides very strong constraints. Only two possibilities can be found at each end of the spin chain: one is the case studied  in \cite{Bai}, and the other is without any boundary interactions. The reflection matrices are anti-diagonal in the first case and proportional to the identity matrix in the second case. Therefore, no more complicated reflection matrices can appear within this class of interactions if we demand integrability.  We construct flavored ABJM theory leading to vanishing boundary interactions at both ends of the open chain through planar two-loop ADM in the scalar sector  by including scalar fields in the fundamental representation and certain six-scalar  interactions,. This flavored ABJM theory is non-supersymmetric.

We will provide the setup in the next section by introducing the spin chain and the Hamiltonian. Section 3 is the main part of this paper, where the coordinate Bethe ansatz is used to determine the integrable spin chain among the class in Section 2. In Section 4, the non-supersymmetric flavored ABJM theory leading to trivial boundary interactions is constructed. Finally, Section 5 presents the conclusion and discussions.

\section{Setup}
The bulk part of the open chain under investigation is of  an alternating type. The bulk interactions are invariant under group $SU(4)$, and  the spin on the odd (even) site is in the $\bf{4}$ ($\bf{\bar{4}}$)
representation of  this group. The boundary interaction terms break $SU(4)$ into $SU(2)\times SU(2)$ and
under this unbroken subgroup, the spin on the odd (even) site is in the $(\bf{2}, \bf{2})$ $((\bf{\bar{2}}, \bf{\bar{2}}))$ representation. The spin on the left (right) boundary
site is in the $(\bf{\bar{2}}, \bf{1})$ ($(\bf{2}, \bf{1})$) representation. Using the notations of \cite{Bai}, the state on this chain can be denoted as follows:
\be |Y^{\dg}_iX^{i_1A_1}X^{\dg}_{i_2A_2}\cdots X^{i_{2L-1}A_{2L-1}}X^{\dg}_{i_{2L}A_{2L}}Y^{i^\prime} \rangle.\ee
Here $X^{i_{2l-1}A_{2l-1}}$ denotes the spin on the odd site which is in the $(\bf{2}, \bf{2})$ representation, $X^\dg_{i_{2l}A_{2l}}$ denotes the spin on the even site which is in the $(\bf{\bar{2}}, \bf{\bar{2}})$ representation. $Y^\dg_i$ denotes the spin in the left boundary which is on the $({\bf{\bar 2}}, \bf{1})$ representation and $Y^{i^\prime}$ denotes the spin on the right boundary which is in the $(\bf{2}, \bf{1})$ representation.
The Hamiltonian acting on the spin chain is chosen as follows:
\bea
{\ml{H}}={\ml{H}}_l+{\ml{H}}_r+{\ml{H}}_{bulk},\label{H1}
\eea
with
\bea
\left({\ml{H}}_l\right)^{j, i_1A_1, j_2B_2}_{i, j_1B_1, i_2A_2}=\left(-a\de^{A_1}_{B_1}\de^{B_2}_{A_2}+b\de^{A_1}_{A_2}\de^{B_2}_{B_1}\right)\cdot\de^{j_2}_i\de^{i_1}_{j_1}\de^j_{i_2}
+c\de^{A_1}_{B_1}\de^{B_2}_{A_2}\cdot\de^{i_1}_i\de^j_{j_1}\de^{j_2}_{i_2},
\eea
\bea
\left({\ml{H}}_r\right)^{i_{2L-1}A_{2L-1}, j_{2L}B_{2L}, i'}_{j_{2L-1}B_{2L-1}, i_{2L}A_{2L}, j'}
&=&\left(-a^\prime\de^{B_{2L}}_{A_{2L}}\de^{A_{2L-1}}_{B_{2L-1}}+b^\prime\de^{B_{2L}}_{B_{2L-1}}\de^{A_{2L-1}}_{A_{2L}}\right)
\cdot\de^{i_{2L-1}}_{j'}\de^{j_{2L}}_{i_{2L}}\de^{i'}_{j_{2L-1}}
\\\no
&+&c^\prime\de^{B_{2L}}_{A_{2L}}\de^{A_{2L-1}}_{B_{2L-1}}\cdot\de^{i_{2L-1}}_{j_{2L-1}}\de^{i'}_{i_{2L}}\de^{j_{2L}}_{j'},
\eea
\bea
{\ml{H}}_{bulk}=\sum_{l=1}^{2L-2}\left(\mathbb{I}_{l,l+1}-\mathbb{P}_{l,l+2}+\frac{1}{2}\mathbb{P}_{l,l+2}\mathbb{K}_{l,l+1}+\frac{1}{2}\mathbb{K}_{l,l+1}\mathbb{P}_{l,l+2}\right),
\eea
where the basic operators $\mathbb{I}$, $\mathbb{P}$, and $\mathbb{K}$ are defined as follows:
\bea
\left(\mathbb{I}_{l,l+1}\right)^{iA,\,j'B'}_{jB,\,i'A'}=\de^i_j\de^{j'}_{i'}\de^A_B\de^{B'}_{A'},\quad \left(\mathbb{P}_{l,l+2}\right)^{iA,\,i'A'}_{jB,\,j'B'}=\de^i_{j'}\de^{i'}_j\de^A_{B'}\de^{A'}_B,\quad
\left(\mathbb{K}_{l,l+1}\right)^{iA,\,j'B'}_{jB,\,i'A'}=\de^i_{i'}\de^{j'}_j\de^A_{A'}\de^{B'}_B.
\eea

Compared with the Hamiltonian in \cite{Bai}, we generalized the coefficients of the non-trivial boundary interaction terms to  arbitrary constants\footnote{An overall factor for the whole Hamiltonian and
 a term proportional to the identity operator are omitted because these will not affect the integrability of  the Hamiltonian. }.
 This study aims  to determine the values of $a, b, c, a^\prime, b^\prime, c^\prime$, which can lead to an integrable open chain.

 \section{Coordinate Bethe ansatz}
 The main computation tool here is the coordinate Bethe ansatz. We first relabel $X^{iA}$ as follows:
 \be (X^{11}, X^{12}, X^{21}, X^{22})=(A_1, A_2, B_1^\dagger, B_2^\dagger). \ee
 The vacuum state is chosen as \be |vac\rangle=|Y_2^\dag (A_1B_2)\cdots (A_1B_2) Y^1\rangle.\ee
The vacuum state is an eigenstate of the Hamiltonian with eigenvalue $-a-a^\prime$.

The one-particle excitation states have three types:
\bea
\mbox{bulk A type}: \quad |Y^{\dg}_2(A_1B_2)\cdots(A_2B_2)\cdots(A_1B_2)Y^1\rangle,&&\\
|Y^{\dg}_2(A_1B_2)\cdots(B^{\dg}_1B_2)\cdots(A_1B_2)Y^1\rangle,&&\\
\mbox{bulk B type}: \quad |Y^{\dg}_2(A_1B_2)\cdots(A_1A^{\dg}_2)\cdots(A_1B_2)Y^1\rangle,&&\\
|Y^{\dg}_2(A_1B_2)\cdots(A_1B_1)\cdots(A_1B_2)Y^1\rangle,&&\\
\mbox{boundary\quad}: \quad |Y^{\dg}_1(A_1B_2)\cdots(A_1B_2)\cdots(A_1B_2)Y^1\rangle,&&\\
|Y^{\dg}_2(A_1B_2)\cdots(A_1B_2)\cdots(A_1B_2)Y^2\rangle.&&
\eea
By bulk A(B) type, we mean a one-particle excitation at an odd(even) side on the bulk part of the chain.
The bulk one-particle will be labeled by the position and excitation type.  A-type excitations on the $(2x-1)$-th bulk site will be denoted
by $|x\rangle_{A_2}$ and $|x\rangle_{B^\dg_1}$, while the B-type excitations on the $2x$-th bulk site will be denoted as $|x\rangle_{A^\dg_2}$ and $|x\rangle_{B_1}$. By $|x\rangle$, we mean one of these four bulk one-particle excitations. The two boundary excitations will be denoted
as $|l\rangle_{Y^\dg_1}$ and $|r\rangle_{Y^2}$.

The bulk two-body $S$-matrix was computed in \cite{Ahn:2009zg}.
We list herein the result with the convention being  that in  $S_{I_1I_2}^{J_1J_2}$,
$I_i$ is used to denote the in-state of the $i$-th particle and $J_i$ is for the out-state of the $i$-th particle.

We define \be u_i\equiv \frac12\cot\frac{k_i}2, \quad u_{ij}\equiv u_i-u_j. \ee
The non-zero elements of the bulk $S$-matrix is
\bea S^{\phi\phi}_{\phi\phi}(k_2, k_1)=\frac{u_{21}+i}{u_{21}-i}, \eea
where $\phi$ is one of $A_2, B_1^\dagger, A_2^\dagger, B_1$;
\bea S^{A_2B_1^\dg}_{A_2B_1^\dg}(k_2, k_1)=S^{B_1^\dg A_2}_{B_1^\dg A_2}(k_2, k_1)=S^{A_2^\dg B_1}_{A_2^\dg B_1}(k_2, k_1)=
S^{B_1A_2^\dg}_{B_1A_2^\dg }(k_2, k_1)=\frac{u_{21}}{u_{21}-i}; \eea
\bea S^{B_1^\dg A_2}_{A_2B_1^\dg}(k_2, k_1)=S^{A_2B_1^\dg }_{B_1^\dg A_2}(k_2, k_1)=S^{B_1A_2^\dg }_{A_2^\dg B_1}(k_2, k_1)=
S^{A_2^\dg B_1}_{B_1A_2^\dg }(k_2, k_1)=\frac{i}{u_{21}-i}; \eea
\bea  S^{A_2B_1}_{A_2B_1}(k_2, k_1)=S^{B_1 A_2}_{B_1 A_2}(k_2, k_1)=S^{A_2^\dg B_1^\dg}_{A_2^\dg B_1^\dg}(k_2, k_1)=
S^{B_1^\dg A_2^\dg}_{B_1^\dg A_2^\dg }(k_2, k_1)=1; \eea

\bea S^{A_2A_2^\dg}_{A_2A_2^\dg}(k_2, k_1)=S^{B_1^\dg B_1}_{B_1^\dg B_1}(k_2, k_1)=S^{A_2^\dg A_2}_{A_2^\dg A_2}(k_2, k_1)=
S^{B_1B_1^\dg}_{B_1B_1^\dg }(k_2, k_1)=\frac{u_{12}}{u_{12}-i}; \eea

\bea S^{A_2A_2^\dg}_{B_1^\dg B_1}(k_2, k_1)=S^{B_1^\dg B_1}_{A_2A_2^\dg}(k_2, k_1)=S^{A_2^\dg A_2}_{B_1B_1^\dg}(k_2, k_1)=
S^{B_1B_1^\dg}_{A_2^\dg A_2}(k_2, k_1)=\frac{i}{u_{12}-i}. \eea

 \cite{Bai} checked that this S-matrix satisfied the Yang--Baxter equations (YBE). Hence, to obtain an integrable open chain, we only need to find the reflection matrices and confirm that these matrices, together with the bulk S-matrix, satisfy the reflection equation. To compute the reflection matrices, we first computed the action of the Hamiltonian on the one-particle state. The computations were straightforward and we listed the results as follows:
\bea {\ml{H}}_{bulk}|x\rangle&=&2|x\rangle-|x+1\rangle-|x-1\rangle, 2\leq x\leq L-1, \label{Hbulk}\\
{\ml{H}}_{bulk}|1\rangle&=&|1\rangle-|2\rangle, \\
{\ml{H}}_{bulk}|L\rangle&=&|L\rangle-|L-1\rangle,\\
{\ml{H}}_{bulk}|l\rangle_{Y_1^\dag}&=&{\ml{H}}_{bulk}|r\rangle_{Y^2}=0,\eea
\bea
{\ml{H}}_l|x\rangle&=&-a|x\rangle, x>1,\\
{\ml{H}}_{l}|r\rangle_{Y^2}&=&-a|r\rangle_{Y^2},\\
{\ml{H}}_l|1\rangle_{A_2}&=&(b-a)|1\rangle_{A_2}+b|1\rangle_{B_1},\\
{\ml{H}}_l|1\rangle_{B_1}&=&(b-a)|1\rangle_{B_1}+b|1\rangle_{A_2},\\
{\ml{H}}_l|1\rangle_{B_1^\dag}&=&(c-a)|1\rangle_{B_1^\dag}+c|l\rangle_{Y^\dag_1},\\
{\ml{H}}_l|1\rangle_{A^\dag_2}&=&-a|l\rangle_{Y^\dg_1},\\
{\ml{H}}_l|l\rangle_{Y^\dg_1}&=&-a|1\rangle_{A^\dg_2}+c|l\rangle_{Y^\dg_1}+c|1\rangle_{B_1^\dg},\eea
\bea
{\ml{H}}_r|x\rangle&=&-a^\prime|x\rangle, x<L,\\
{\ml{H}}_r|l\rangle_{Y^\dg_1}&=&-a^\prime|l\rangle_{Y^\dg_1},\\
{\ml{H}}_r|L\rangle_{A_2}&=&(b^\prime-a^\prime)|L\rangle_{A_2}+b^\prime |L\rangle_{B_1},\\
{\ml{H}}_r|L\rangle_{B_1}&=&(b^\prime-a^\prime)|L\rangle_{B_1}+b^\prime |L\rangle_{A_2},\\
{\ml{H}}_r|L\rangle_{B_1^\dag}&=&-a^\prime|r\rangle_{Y^2},\\
{\ml{H}}_r|L\rangle_{A_2^\dag}&=&(c^\prime-a^\prime)|L\rangle_{A_2^\dag}+c^\prime |r\rangle_{Y^2},\\
{\ml{H}}_r|r\rangle_{Y^2}&=&-a^\prime |L\rangle_{B_1^\dag}+c^\prime|L\rangle_{A_2^\dag}+c^\prime|r\rangle_{Y^2}.\label{Hright}
\eea
We want to find the spin wave eigenstate of ${\cal H}$. Thus, we need to consider the superposition of $|x\rangle_{A_2}$ and $|x\rangle_{B_1}$
and the superposition of $|x\rangle_{A_2^\dag}$, $|x\rangle_{B_1^\dg}$, $|l\rangle_{Y^\dg_1}$ and $|r\rangle_{Y^2}$. Let us start with the ansatz involving $|x\rangle_{A_2}$ and $|x\rangle_{B_1}$,
\be |\psi_1(k)\rangle=\sum_{x=1}^L f(x)|x\rangle_{A_2}+g(x)|x\rangle_{B_1}, \ee
with \bea f(x)&=&Fe^{ikx}+\tilde{F}e^{-ikx},\label{fwave}\\
          g(x)&=&Ge^{ikx}+\tilde{G}e^{-ikx}. \label{gwave} \eea
We can obtain the following from  Eqs.~(\ref{Hbulk}--\ref{Hright}):
\bea  {\cal H}|\psi_1(k)\rangle&=&\sum_{x=2}^{L-1} [(2-a-a^\prime)f(x)-f(x-1)-f(x+1)]|x\rangle_{A_2}\nonumber\\
                               &+&\sum_{x=2}^{L-1} [(2-a-a^\prime)g(x)-g(x-1)-g(x+1)]|x\rangle_{B_1}\nonumber\\
                               &+&[(1+b-a-a^\prime)f(1)-f(2)+bg(1)]|1\rangle_{A_2}\nonumber\\
                               &+&[(1+b-a-a^\prime)g(1)-g(2)+bf(1)]|1\rangle_{B_1} \nonumber\\
                               &+&[(1+b^\prime-a-a^\prime)f(L)-f(L-1)+b^\prime g(L)]|L\rangle_{A_2}\nonumber\\
                               &+&[(1+b^\prime-a-a^\prime)g(L)-g(L-1)+b^\prime f(L)]|L\rangle_{B_1}.
\eea

The eigenvalue equation ${\cal H}|\psi_1(k)\rangle=E(k)|\psi_1(k)\rangle$ then provides:
\bea (2-a-a^\prime)f(x)-f(x+1)-f(x-1)&=&E(k) f(x), 2\leq x\leq L-1, \label{fx}\\
     (2-a-a^\prime)g(x)-g(x+1)-g(x-1)&=&E(k) g(x), 2\leq x\leq L-1, \label{gx}\\
     (1+b-a-a^\prime)f(1)-f(2)+bg(1)&=&E(k) f(1), \label{f1}\\
     (1+b-a-a^\prime)g(1)-g(2)+bf(1)&=&E(k) g(1), \label{g1}\\
     (1+b^\prime-a-a^\prime)f(L)-f(L-1)+b^\prime g(L)&=&E(k) f(L), \label{fL}\\
     (1+b^\prime-a-a^\prime)g(L)-g(L-1)+b^\prime f(L)&=&E(k) g(L), \label{gL}.
     \eea

Eqs.~(\ref{fwave}) and (\ref{fx}) provide the following dispersion relation:
\be E(k)=2-2\cos k-a-a^\prime,\label{dr} \ee
which is the same as the results from Eqs.~(\ref{gwave}) and (\ref{gx}). We can obtain the following from Eqs.~(\ref{f1}--\ref{g1}) and the dispersion relation Eq.~(\ref{dr}):
\be \left(\begin{array}{c}F\\
  G\end{array}\right)= K_{l1}(k)\left(\begin{array}{c}\tilde{F}\\
  \tilde{G}\end{array}\right),\ee
with the left reflection matrix in this sector $K_{l1}(k)$ being,
\bea K_{l1}(k)=\left(\begin{array}{cc}
ln_{11}/ld_1&ln_{12}/ld_1\\
ln_{12}/ld_1&ln_{11}/ld_1
\end{array}\right),\eea
with \bea ln_{11}&=&(b-1)(e^{ik}-1),\\
ln_{12}&=&-b(e^{ik}+1),\\
ld_1&=&e^{ik}+(2b-1)e^{2ik} \eea
Eqs.~(\ref{fL}) and (\ref{gL}), with the dispersion relation Eq.~(\ref{dr}),  will provide the right reflection matrix in this sector.
We used the definition of the right reflection matrix in  \cite{Bai}, which was equivalent to the definition in \cite{Li:2013vdh}.
\bea
 e^{2ikL}\left(\begin{array}{c}F\\
  G\end{array}\right)\equiv K_{r1}(k)\left(\begin{array}{c}\tilde{F}\\
  \tilde{G}\end{array}\right).\eea
 The result is
 \bea K_{r1}(k)=\left(\begin{array}{cc}rn_{11}/rd_1&rn_{12}/rd_1\\
rn_{12}/rd_1&rn_{11}/rd_1, \end{array}\right) \eea
with \bea  rn_{11}&=&-e^{-ik}(e^{ik}-1)(b^\prime-1),\\
rn_{12}&=&-e^{-ik}b^\prime(e^{ik}+1),\\
rd_1&=&e^{ik}-1+2b^\prime.
\eea

Let us now turn to the section involving $|x\rangle_{B_1^\dag}, |x\rangle_{A_2^\dag}, |l\rangle_{Y_1^\dag}$ and $|r\rangle_{Y^2}$ by starting with the
ansatz,
\bea|\psi_2(k)\rangle=\sum_{x=1}^L(n(x)|x\rangle_{B_1^\dag}+h(x)|x\rangle_{A_2^\dag})+\beta|l\rangle_{Y^\dag_1}+\gamma|r\rangle_{Y^2},\eea
with \bea n(x)=Ne^{ikx}+\tilde{N}e^{-ikx},\label{nwave}\\
          h(x)=He^{ikx}+\tilde{H}e^{-ikx}. \label{hwave}  \eea
The eigenvalue equation ${\cal H}|\psi_2(k)\rangle=E(k)|\psi_2(k)$ provides:
\bea (2-a-a^\prime)n(x)-n(x-1)-n(x+1)&=&E(k)n(x), \label{nx}\\
      (2-a-a^\prime)h(x)-h(x-1)-h(x+1)&=&E(k)h(x), \label{hx}\\
      (1+c-a-a^\prime)n(1)-n(2)+c\beta&=&E(k)n(1), \label{n1}\\
      (1-a^\prime)h(1)-h(2)-a\beta&=&E(k)h(1) \label{h1}\\
      cn(1)-ah(1)+(c-a^\prime)\beta&=&E(k)\beta, \label{beta}\\
      (1-a)n(L)-n(L-1)-a^\prime\gamma&=&E(k)n(L), \label{nL},\\
      (1+c^\prime-a-a^\prime)h(L)-h(L-1)+c^\prime\gamma&=&E(k)h(L), \label{hL},\\
      c^\prime h(L)-a^\prime n(L)+(c^\prime-a)\gamma&=&E(k)\gamma.\label{gamma}
      \eea
Eqs.~(\ref{nwave}) and (\ref{nx}) (Eqs.~(\ref{hwave}) and (\ref{hx})) lead to the same dispersion relation as the one given by Eq.~(\ref{dr}).
With this relation, Eqs.~(\ref{n1}--\ref{beta}) provide:
\bea \left(\begin{array}{c}N\\
  H\end{array}\right)= K_{l2}(k)\left(\begin{array}{c}\tilde{N}\\
  \tilde{H}\end{array}\right),\eea
  with the left reflection matrix in this sector $K_{l2}(k)$ being,
\be K_{l2}(k)=\left(\begin{array}{cc}ln_{21}/ld_2&ln_{22}/ld_2\\
                                     ln_{22}/ld_2&ln_{23}/ld_2 \end{array}\right), \ee
where
\bea ln_{21}&=&(c-1)(1+(2a-3)e^{ik})-(a-1)((2c-3)e^{2ik}+e^{3ik}),
\\
ln_{22}&=&ace^{ik}(e^{ik}+1),\\
ln_{23}&=&(a-1)(1+(2c-3)e^{ik})-(c-1)(e^{3ik}+(2a-3)e^{2ik}),\\
ld_2&=&-e^{ik}+(3-2a-2c)e^{2ik}+(e^{4ik}-3e^{3ik})(a-1)(c-1).\eea
Similarly, Eqs.~(\ref{nL}--\ref{gamma}) provide the right  reflection matrix in this sector $K_{r2}(k)$ defined by:
\bea
 e^{2ikL}\left(\begin{array}{c}N\\
  H\end{array}\right)\equiv K_{r2}(k)\left(\begin{array}{c}\tilde{N}\\
  \tilde{H}\end{array}\right),\eea
as
\bea K_{r2}=\left(\begin{array}{cc}rn_{21}/rd_2&rn_{22}/rd_2\\
rn_{22}/rd_2&rn_{23}/rd_2\end{array}\right),\eea
with
\bea rn_{21}&=&e^{-ik}(c^\prime-1)+(2a^\prime-3)(c^\prime-1)+e^{ik}(a^\prime-1)(2c^\prime-3)-e^{2ik}(a^\prime-1),\\
     rn_{22}&=&-a^\prime c^\prime(1+e^{ik}),\\
     rn_{23}&=&e^{-ik}(a^\prime-1)+(2c^\prime-3)(a^\prime-1)+e^{ik}(c^\prime-1)(2a^\prime-3)-e^{2ik}(c^\prime-1),\\
     rd_2&=&e^{3ik}+(3e^{ik}-1)(a^\prime-1)(c^\prime-1)+e^{2ik}(2a^\prime+2c^\prime-3). \eea
The full left  reflection matrix is presented as follows:
\be K_l(k)=\left(\begin{array}{cccc}
ln_{11}/ld_1&0&0&ln_{12}/ld_1\\
0&ln_{21}/ld_2&ln_{22}/ld_2&0\\
0&ln_{22}/ld_2&ln_{23}/ld_2&0 \\
ln_{12}/ld_1&0&0&ln_{11}/ld_1
\end{array}\right) ,\ee
while the full right reflection matrix is:
\be K_r(k)=\left(\begin{array}{cccc}
rn_{11}/rd_1&0&0&rn_{12}/rd_1\\
0&rn_{21}/rd_2&rn_{22}/rd_2&0\\
0&rn_{22}/rd_2&rn_{23}/rd_2&0 \\
rn_{12}/rd_1&0&0&rn_{11}/rd_1
\end{array}\right) ,\ee
where the order of the excitations has already been chosen as $A_2, B_1^\dag, A_2^\dag, B_1$.

With the help of \texttt{Mathematica}, we found the solutions of the left--boundary Yang--Baxter equation \cite{Cherednik:1985vs}\footnote{The boundary
Yang--Baxter equations in the algebraic Bethe ansatz approach were first presented in \cite{Sklyanin:1988yz}.} to be:
\bea
&\left[S(k_1,k_2)\right]^{m_1m_2}_{l_1l_2}\left[K_l(k_2)\right]^{l_2}_{j_2}\left[S(-k_2,k_1)\right]^{l_1j_2}_{j_1i_2}\left[K_l(k_1)\right]^{j_1}_{i_1}\\\no
&=\left[K_l(k_1)\right]^{m_1}_{l_1}\left[S(-k_1,k_2)\right]^{l_1m_2}_{j_1l_2}\left[K_{l}(k_2)\right]^{l_2}_{j_2}\left[S(-k_2,-k_1)\right]^{j_1j_2}_{i_1i_2},
\eea
are $a=b=c=c_l$ with $c_l=0$ or $1$.
The first case provides a reflection matrix proportional to the identity matrix,
\be  K_l(k)=e^{ik}I_{4\times 4},\ee
while the second case is  the one studied in \cite{Bai} with the reflection matrix:
  \bea
  K_l(k)=
  \left
  (\begin{array}{cccc}0&0&0&-e^{-ik}\\0&0&-1&0\\
  0&-1&0&0\\
  -e^{-ik}&0&0&0
  \end{array}
  \right),
  \eea
 Similar to the solutions of the right--boundary Yang--Baxter equation
\bea
&\left[S(-k_1,-k_2)\right]^{m_1m_2}_{l_1l_2}\left[K_r(-k_1)\right]^{l_1}_{j_1}\left[S(-k_2,k_1)\right]^{j_1l_2}_{i_1j_2}\left[K_{r}(-k_2)\right]^{j_2}_{i_2}\\\no
&=\left[K_r(-k_2)\right]^{m_2}_{l_2}\left[S(-k_1,k_2)\right]^{m_1l_2}_{l_1j_2}\left[K_r(-k_1)\right]^{l_1}_{j_1}\left[S(k_2,k_1)\right]^{j_1j_2}_{i_1i_2},
\eea
are $a^\prime=b^\prime=c^\prime=c_r$ with $c_r=0$ or $1$,
with the reflection matrix
\be  K_r(k)=e^{ik}I_{4\times 4},\ee
or
     \bea
  K_r(k)=
  \left
  (\begin{array}{cccc}0&0&0&-e^{-ik}\\0&0&-e^{-2ik}&0\\
  0&-e^{-2ik}&0&0\\
  -e^{-ik}&0&0&0
  \end{array}
  \right).
  \eea
The latter case is the one studied in \cite{Bai}.

\section{Flavored ABJM theory for the new integrable boundary interactions}
In this section, we briefly construct planar flavored ABJM theory which leads to vanishing boundary interactions at both ends of the open chain (i. e., $c_l=c_r=0$) from the two-loop ADM in the scalar sector.
 Notice the bulk two-site trace operators near the boundaries from the gauge and fermion interactions \cite{Minahan:2008, Bak:2008}  that should be cancelled by the boundary contributions from the two-loop calculations. Accordingly we add to the ABJM theory scalars $Y^i, i=1, 2$ in the $({\bf{N_c}}, \bf{1})$ representation of the gauge group $U(N_c)\times U(N_c)$, then add the following terms to the ABJM action,\footnote{Considering the case with $N_f=1$ is enough, and we recall that $X^{iA}$ is in the $({\bf N_c}, {\bf \bar{N}_c})$ representation of the gauge group $U(N_c)\times U(N_c)$.}
 \bea
 \ml{S}_{fund.}&=&\int d^3x \, \mbox{tr}\left[-\mathcal{D}_{\mu}Y^{\dagger}_i\mathcal{D}^{\mu}Y^i+\frac{4\pi^2}{k^2} X^{iA}X^\dagger_{iA}
 X^{jB}X^\dagger_{jB}Y^kY^\dagger_k\right],\label{fund}
  \eea
  where,
  \be \mathcal{D}_\mu Y^a=\partial_\mu Y^a+i A_\mu Y^a, \ee
with $A_\mu$ being the gauge field in the adjoint representation of the first $U(N_c)$.
The computations in \cite{Bai} showed that the abovementioned interaction will contribute  the following terms in the spin chain Hamiltonian
\bea
{\ml{H}}^{j, i_1A_1, j_2B_2}_{i, j_1B_1, i_2A_2}=-\frac{\lambda^2}{4}\delta^j_i\delta^{i_1}_{i_2}\delta^{j_2}_{j_1}\delta^{A_1}_{A_2}
\delta^{B_2}_{B_1},
\eea
and
\bea
{\ml{H}}^{i_{2L-1}A_{2L-1}, j_{2L}B_{2L}, i'}_{j_{2L-1}B_{2L-1}, i_{2L}A_{2L}, j'}=-\frac{\lambda^2}{4}\delta^{i'}_{j'}\delta^{j_{2L}}_{j_{2L-1}}\delta^{i_{2L-1}}_{i_{2L}}
\delta^{A_{2L-1}}_{A_{2L}}\delta_{B_{2L-1}}^{B_{2L}},
\eea
in the 't Hooft limit.
These terms
 exactly cancel  the abovementioned bulk two-site trace operators near the boundaries. Hence, the flavored ABJM theory obtained from adding (\ref{fund})
 to the ABJM action exactly provides the spin chain Hamiltonian with $c_l=c_r=0$. We do not include fermions in the $(\bf{N_c}, \bf{1})$ representation of the gauge group, hence, this flavored ABJM theory is non-supersymmetric.

\section{Conclusion and discussions}
The two-loop ADM in the scalar sector of planar ABJM theory provides an integrable Hamiltonian on a closed alternating spin chain \cite{Minahan:2008, Bak:2008}. Adding flavors to this theory  provides an integrable open spin chain with non-trivial boundary interactions and anti-diagonal boundary reflection matrices \cite{Bai}. In this study, we considered whether one can obtain more complicated reflection matrices that still satisfy the boundary Yang--Baxter equations  by changing the coefficients in the  boundary interactions. We found that the only other solution at each end of the spin chain is with vanishing boundary interactions. Moreover, the corresponding reflection matrices
are proportional to the identity matrix. The flavored ABJM theory providing  the vanishing boundary interactions was constructed and it is non-supersymmetric.
Generalizing this study to more general boundary interactions, including those leading to  other patterns of breaking $SU(4)$ symmetry and to the higher rank cases, will be interesting. We leave these to future studies.

\section*{Acknowledgments}
JW would like to thank Weng-Li Yang for very helpful discussions.
This work was in part supported by National Natural Science Foundation of China under Grant No. 11575202. JW would like to thank ICTS-USTC for hospitality where this work is presented. JW would like to thank the participants of the advanced workshop ``Dark Energy and Fundamental Theory" supported by the Special Fund for Theoretical Physics from the National Natural Science Foundations of China with Grant No. 11447613 for stimulating discussion.

\end{document}